\documentclass{llncs}
\usepackage[T1]{fontenc}
\usepackage[utf8]{inputenc}
\usepackage[english]{babel}
\usepackage{cite}
\usepackage{subfig}
\usepackage{graphicx}
\usepackage{booktabs}
\usepackage{multirow}
\usepackage[pdftex]{hyperref}
\begin{document}

\title{assistME: A Platform for Assisting Engineers in Maintaining the Factory Pipeline}

\author{Ragaad AlTarawneh\inst{1}, Jens Bauer\inst{1}, Nicole Menck\inst{2}, Shah Rukh Humayoun\inst{1}, Achim Ebert\inst{1} }
\institute{Computer Graphics and HCI Group \and		
Institute for Manufacturing Technology and Production Systems (FBK) \\
 University of Kaiserslautern, Germany\\
  \inst{1}\email{ \{tarawneh,j\textunderscore bauer,humayoun,ebert\}@cs.uni-kl.de}\\ 
  \inst{2}\email{ nicole.menck@mv.uni-kl.de}
 }


\maketitle

\begin{abstract}
In this position paper, we present our approach of utilizing mobile devices (i.e., mobile phones and tablets) for assisting engineers and experts in understanding and maintaining the factory pipelines. For this, we present a platform, called assistME, that is composed of three main components: the assistME Server, the assistME mobile infrastructure, and the co-assistME collaborative environment. In order to get full utilization of the assistME platform, we assume that an initial setup is made in the factory in such a way that it is equipped with different sensors to collect data about specific events in the factory pipeline together with the corresponding locations of these events. The assistME Server works as a central control unit in the platform and collects data from the installed sensors in the factory pipeline. In the case of any unexpected behavior or any critical situation in the factory pipeline, notification and other details are sent to the related group of engineers and experts through the assistME mobile app. Further, the co-assistME collaborative environment, equipped with a large shared screen and multiple mobile devices, helps the engineers and experts to collaborate with to understand and analyze the current situation in the factory pipeline in order to maintain it accurately.  

\end{abstract}

\section{Introduction}
\label{sec:Introduction}
%
In the application presented in this paper, we focus on helping engineers and experts in collaborating to understand different aspects of heterogeneous systems in order to maintain them accurately and in a timely manner. Nowadays, heterogeneous systems like embedded systems (i.e., smart cards, cars, washing machines, robots, airplanes, etc.) are involved in many activities of our modern daily life~\cite{embeddedsystem}.

The complexity level of these systems usually is higher due to the fact that they are composed of different sub-systems with different types (either hardware, software, or a set of interfaces). This makes the process of analyzing and maintaining these systems relatively a difficult task~\cite{Kaiser_Liggesmeyer2003}. However, because of their usage in performing many of our daily life tasks it is critical to understand different aspects about the underlying system as well as those situations that could cause the failure of the whole system. However, this understanding requires an intensive collaboration between different engineers and experts who design, implement, and maintain them.

Involvement of different domain experts (e.g. engineers) having different backgrounds in the analysis and maintaining process means a difference not only in the used terms (domain specific language) to describe the system status but also a difference in seeing the system from their own perspective\cite{AlTarawneh:CSCW2014}. For example, \textit{safety experts} are responsible for analyzing the failure relations between the system components while \textit{system engineers} are more interested in understanding the structural relations between the system components. Therefore, a platform is required for unifying the differences and bringing the participated parties (e.g., system engineers and safety experts) together for a better and accurate understanding of different aspect (e.g., the safety and the reliability aspects) of these complex systems\cite{ AlTarawneh:IUI2014 }.

In the above context of analyzing and maintaining heterogeneous systems, our application area is an intelligent-factory, which supports the collection of real data about its components in real time using a set of sensors distributed over the factory pipeline. We assume that in such factory each group of sensors are attached to a specific group of components in the pipeline, where these sensors collect data regarding these components' behavior during the production process. Further, we assume that this collected data is periodically sent to some central server in order to utilize for analysis and maintaining purpose.

To support the analysis and maintaining process of such a set of heterogeneous systems (i.e., some intelligent factory), we propose a platform, called \textbf{assistME}, that utilizes current smart mobile devices and mobile technology as well as a collaborative environment. In the remainder of the paper, we focus on our assistME platform and its different components.


\section{The assistME Platform Overview}
The current mobile devices and technologies play an important role in the infrastructure of the assistME platform in order to provide engineers and experts an environment where they can analyze the system status (i.e., in our application area a factory pipeline) and then can collaborate for maintaining it more efficiently. The proposed assistME platform is mainly composed of three components: the assistME Server, the assistME mobile infrastructure, and the co-assistME collaborative environment. 

The assistME Server works as a central control unit in the platform and is connected with the installed sensors in the factory in order to collect data from these sensors. This data is then analyzed and in the case of an unexpected behavior in the factory pipeline or any critical situation, a notification about the current situation of the factory status is sent by assistME server to the responsible engineers and experts through the assistME mobile infrastructure. The engineers and experts can also analyze the underlying system from different aspects (e.g., an abstract graph representation for showing the failure information or the system 3D model for showing the system actual model) on the mobile devices on-demand.   

Further, the co-assistME collaborative environment, equipped with a large shared screen and mobile devices, helps the engineers and experts to collaborate with each other in order to understand more accurately the system status and to decide the operations for maintaining the factory pipeline. In this case, engineers and experts use co-assistME mobile app for interacting with the large shared screen in the provided collaborative environment as well as for sharing information with each other. 

\begin{figure}[!htb]
\centering
\includegraphics[width=0.99\textwidth, totalheight=0.5\textheight]{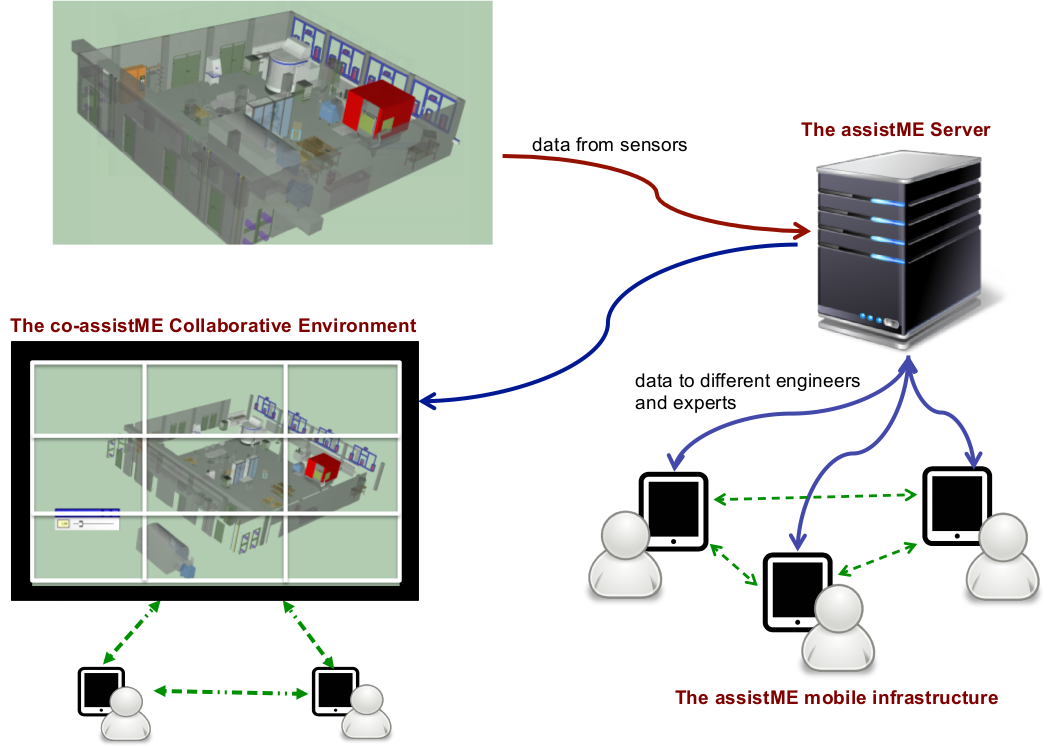}
\caption{The assistME Platform Overview}  
\label{fig:assistMeSetUp}
\end{figure}

Figure~\ref{fig:assistMeSetUp} provides an overview of the whole assistME platform. In the forthcoming sections, first we introduce the assistME mobile infrastructure and then explain the co-assistME collaborative environment.  

\section{The assistME Mobile Infrastructure}

Inspired by the concept of factory of things, introduced by Detlef Zuehlke in~\cite{Zuehlke10arc}, we designed the assistME mobile infrastructure in a way that supports the communication and collaboration between different teams (composed of related engineers and experts) in dynamic and complex environments, like the intelligent factories in our case. One of the main components of our assistME mobile infrastructure is a mobile app that informs (after getting notification from the assistME server) the current situation of the factory status to the responsible engineers and experts and provides synchronization between these engineers and experts in order to support them to update the data between them on demand, which is helpful to resolve the situation more efficiently and effectively. Further, it also enables the engineers and experts to see the system aspects from different perspectives on demand. 

The assistME platform in this regard works as follows: The sensors' data are periodically sent to the assistME server. The assistME server processes the data attributes and measures all the required parameters that are necessary to judge a specific component's status and shows it in the current status report of that system component. The related engineers and experts can ask to view the updated version of the report through the assistME mobile app on demand. In the case of any unexpected behavior of the component (e.g., some parameters' values are not in the expected range) is noted by the assistME server, a notification is sent to the group of related engineers and expert through the assistME mobile app.

In our infrastructure, the smart devices, equipped with assistME mobile app, are connected to the assistME Server in a client-server pattern (see Figure~\ref{fig:assistMeSetUp}). Our current implementation uses the common Internet protocols (e.g. TCP) over Wi-Fi to communicate, but any wireless communication protocol could alternatively be used if security or connectivity is an issue. The infrastructure security can be enhanced by adding application or transport layer encryption. 

\begin{figure}[!htb]
\centering
\includegraphics[width=0.99\textwidth, totalheight=0.5\textheight]{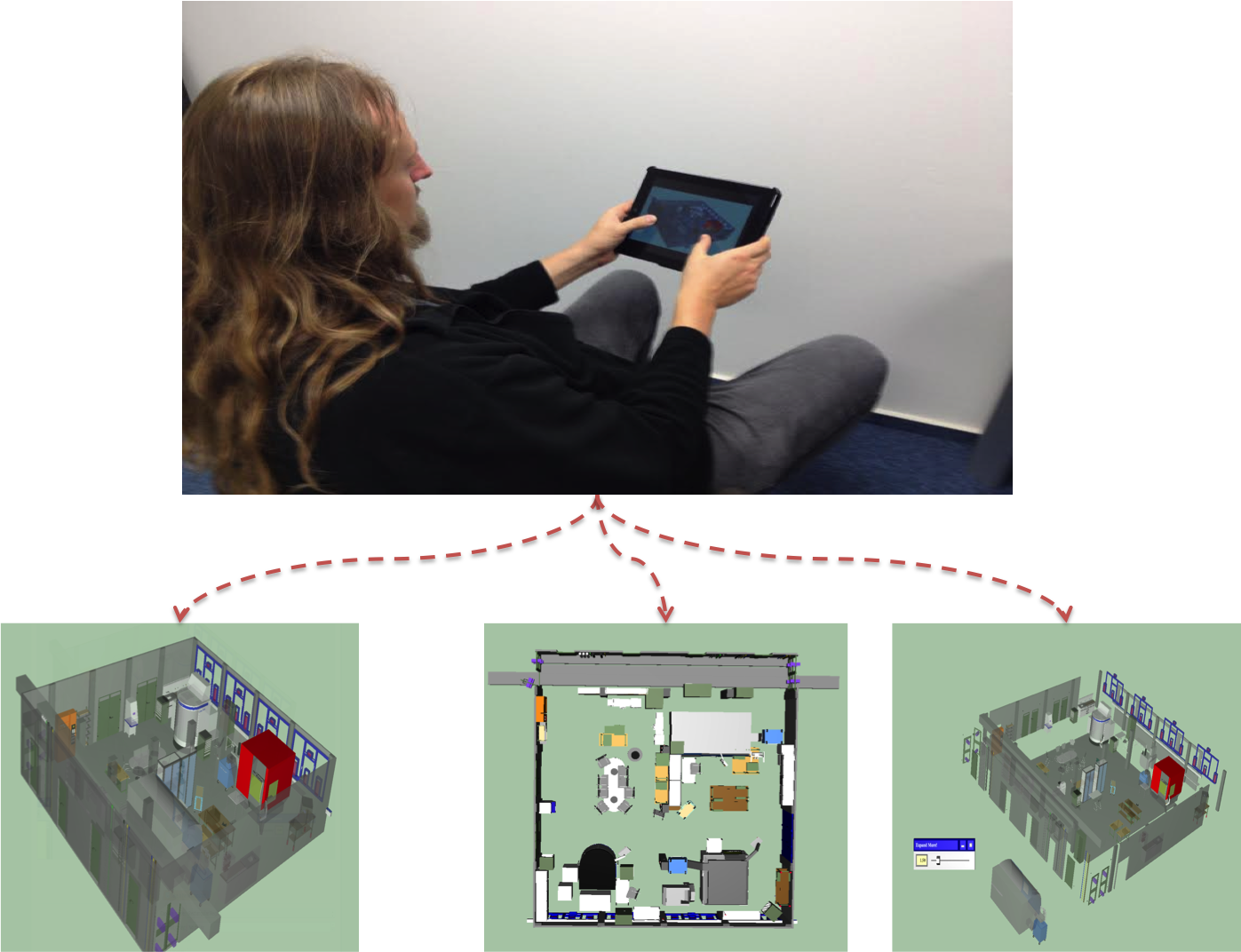}
\caption{An engineer is observing the current status of the factory pipeline on a mobile device, while different possible visualizations of the factory pipeline are shown in the bottom side where the critical components are shown in red color.} 
\label{fig:assistMeoption}
\end{figure}

The assistME platform also offers visualization of the critical components in the factory pipeline on mobile devices (see Figure~\ref{fig:assistMeoption}). The resulting visualization (either of a specific component or a set of components or a complete setup of pipeline) on the mobile device is shown in the 3D form after it is rendered on the assistME Server. On the mobile device, the critical components in the current scenario are shown in red color in order to easily identify them. 

On the mobile device, the assistME mobile app provides different interaction options to explore the received 3D visual representation. For example, it offers the possibility of selecting one of the components in which case the assistME mobile app shows further information about the highlighted component after receiving this information form the assistME Server. Further, engineers and experts can also view the detailed report about the failure scenario and the main parameters values of the selected critical component. Figure~\ref{fig:assistMeoption} provides a screen-shot where an engineer observes the current status of the factory pipeline.

\section{The co-assistME Collaborative Environment}

\begin{figure}[!htb]
\centering
\includegraphics[width=0.99\textwidth, totalheight=0.5\textheight]{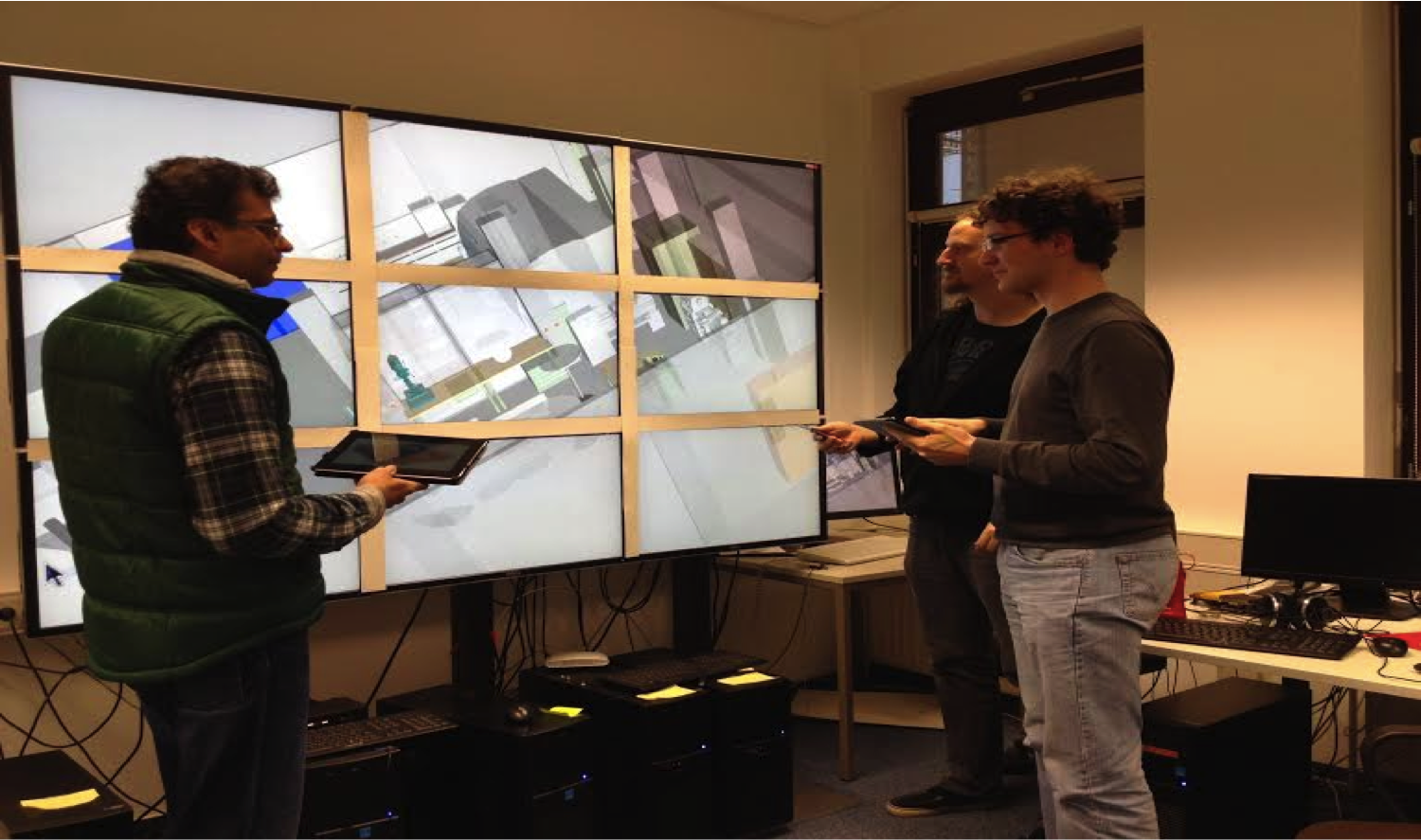}
\caption{The co-assistME collaborative environment, where the large shared screen shows the system structure and critical components while three engineers are using mobile devices (mobilphones and tablets) to interact with it. }  
\label{fig:assistMeCol}
\end{figure}

The co-assistME collaborative environment provides the facility to the interested engineers and experts to collaborate with each other in order to understand accurately the system status or to diagnose the problem so that to maintain the system accordingly. For this, co-assistME collaborative environment is equipped with a large shared screen and mobile devices (see Figure~\ref{fig:assistMeCol}, where the large screen provides the system view and status from different perspectives while the engineers and experts use the mobile devices, equipped with co-assistME mobile app, to interact with the content on the large shared screen as well as to exchange the information with each other.

Our approach of using the mobile devices as input devices to interact with the large shared screen in the collaborative environment is a highly scalable approach, as in this case the input device availability does not affect the number of users the system can handle simultaneously~\cite{Bauer2011}. The only real bottleneck is the bandwidth of the Wi-Fi connection (or other radio network connections). Still, current WiFi standards easily allow more than 50 users to interact simultaneously (not including idling users). Further, the actual available space for users to stand is also an important factor in such environment for the proper collaboration with each other and for the better interaction with the system. 

With the advent of new kinds of smart mobile devices, such as smart watches or data glasses, it is important to mention that these new devices can be used as input devices with our co-assistME collaborative environment without any technical issue. This is especially useful in environments where engineers need their hands free, as then such wearable smart mobile devices can be an interesting and convenient choice.

We are planning to extend our collaborative environment from a single location (we call it \textit{locally collaborative environment}) to multiple locations (we call it \textit{globally collaborative environment}). In globally collaborative environment case, the shared screen at each location would have a view synchronized with all other locations and the reflection on one view at one location due to user interaction with it would also synchronize with other locations. In this case, teams at different locations in a factory would collaborate with each other while staying at their own place.


\section{Conclusion}\label{sec:Conclusion}

In this paper, we presented our assistME platform to support the engineers and experts in understanding and maintaining the factory pipeline. We explained that how we use mobile devices for different purposes, e.g., for notifying the related group of engineers and experts about some unexpected behavior in the factory pipeline or for using them as input device to interact with large shared screen in the collaborative environment. 

In the future, we intend to perform evaluation studies in real environment in order to find out the feasibility and effectiveness of our platform and the approach. Further, we also intend to enhance our collaborative environment to support the globally collaborative environment.
  
\bibliographystyle{abbrv}
\bibliography{llncs}

\begin{thebibliography}{1}

\bibitem{AlTarawneh:CSCW2014}
R.~AlTarawneh, J.~Bauer, and A.~Ebert.
\newblock A visual interactive environment for enhancing collaboration between
  engineers for the safety analysis mechanisms in embedded systems.
\newblock In {\em Proceedings of the Companion Publication of the 17th ACM
  Conference on Computer Supported Cooperative Work 38; Social Computing}, CSCW
  Companion '14, pages 125--128, New York, NY, USA, 2014. ACM.

\bibitem{AlTarawneh:IUI2014}
R.~AlTarawneh, J.~Bauer, S.~R. Humayoun, A.~Ebert, and P.~Liggesmeyer.
\newblock Enhancing understanding of safety aspects in embedded systems through
  an interactive visual tool.
\newblock In {\em Proceedings of the Companion Publication of the 19th
  International Conference on Intelligent User Interfaces}, IUI Companion '14,
  pages 9--12, New York, NY, USA, 2014. ACM.

\bibitem{Bauer2011}
J.~Bauer, S.~Thelen, and A.~Ebert.
\newblock Evaluation of large display interaction using smart phones.
\newblock {\em IEEE Conference on Visual Analytics Science and Technology
  VisWeek 2011}, 2011.

\bibitem{Kaiser_Liggesmeyer2003}
B.~Kaiser, P.~Liggesmeyer, and O.~Mäckel.
\newblock A new component concept for fault trees.
\newblock {\em Reproduction}, 33:37--46, 2003.

\bibitem{embeddedsystem}
E.~A. Lee and S.~A. Seshia.
\newblock {\em Introduction to Embedded Systems - A Cyber-Physical Systems
  Approach}.
\newblock Lee and Seshia, 1 edition, 2010.

\bibitem{Zuehlke10arc}
D.~Zuehlke.
\newblock {SmartFactory}---towards a factory-of-things.
\newblock {\em Annual Reviews in Control}, 34(1):129--138, apr 2010.

\end{thebibliography}

\end{document}